\documentclass[conference]{IEEEtran}
\IEEEoverridecommandlockouts
\usepackage{cite}
\usepackage{amsmath,amssymb,amsfonts}
\usepackage{algorithmic}
\usepackage{graphicx}
\usepackage{textcomp}
\usepackage{xcolor}
\def\BibTeX{{\rm B\kern-.05em{\sc i\kern-.025em b}\kern-.08em
    T\kern-.1667em\lower.7ex\hbox{E}\kern-.125emX}}

\usepackage{acro}
\usepackage{array}
\usepackage{multirow}
\usepackage{caption}
\usepackage{float}
\usepackage{color,soul}
\usepackage{xurl}

\DeclareAcronym{ASiR}{short = ASiR , long = AirScale indoor radiohead}
\DeclareAcronym{BBU}{short = BBU ,  long = baseband unit}
\DeclareAcronym{RAN}{short = RAN , long = radio access network}
\DeclareAcronym{CN}{short = CN , long = core network}
\DeclareAcronym{UE}{short = UE , long = user equipment}
\DeclareAcronym{RE}{short = RE , long = resource element}
\DeclareAcronym{BS}{short = BS , long = base station}
\DeclareAcronym{gNB}{short = gNB , long = next generation NodeB} 
\DeclareAcronym{3GPP}{short = 3GPP , long = Third Generation Partnership Project}
\DeclareAcronym{NR}{short = NR , long = new radio}
\DeclareAcronym{4G}{short = 4G , long = fourth generation}
\DeclareAcronym{5G}{short = 5G , long = fifth generation}
\DeclareAcronym{3G}{short = 3G , long = third generation}
\DeclareAcronym{LTE}{short = LTE , long = long term evolution}
\DeclareAcronym{RF}{short = RF , long = radio frequency}
\DeclareAcronym{RRH}{short = RRH , long = remote radio heads}
\DeclareAcronym{pRRH}{short = pRRH , long = pico-RRH}
\DeclareAcronym{mRRH}{short = mRRH , long = micro-RRH}
\DeclareAcronym{PCI}{short = PCI , long = physical cell ID}
\DeclareAcronym{SSB}{short = SSB , long = synchronization signal block}
\DeclareAcronym{RSS}{short = RSS , long = root sum square}
\DeclareAcronym{DSS}{short = DSS , long = dynamic spectrum sharing}
\DeclareAcronym{TDD}{short = TDD , long = time division duplex}
\DeclareAcronym{SCS}{short = SCS , long = sub-carrier spacing}
\DeclareAcronym{mmWave}{short = mmWave , long = millimeter-wave}

\DeclareAcronym{LOS}{short = LOS , long = line of sight}

\DeclareAcronym{EMF}{short = EMF , long = electromagnetic field}
\DeclareAcronym{SA}{short = SA , long = stand-alone}
\DeclareAcronym{NSA}{short = NSA , long = non-stand-alone}
\DeclareAcronym{ICNIRP}{short = ICNIRP , long = International Commission on Non-Ionizing Radiation Protection}
\DeclareAcronym{EU}{short = EU , long = European Union}
\DeclareAcronym{IEEE}{short = IEEE , long = Institute of Electrical and Electronics Engineers}
\DeclareAcronym{ICES}{short = ICES , long = International Committee on Electromagnetic Safety}
\DeclareAcronym{IEC}{short = IEC , long = International Electrotechnical Commission}
\DeclareAcronym{NATO}{short = NATO , long = North Atlantic Treaty Organization}
\DeclareAcronym{FCC}{short = FCC , long = Federal Communications Commission}
\DeclareAcronym{NCRP}{short = NCRP , long = National Council on Radiation Protection and Measurements}
\DeclareAcronym{WHO}{short = WHO , long = World Health Organization}
\DeclareAcronym{NDAC}{short = NDAC , long = Nokia Digital Automation Cloud}
\DeclareAcronym{TUK}{short = TUK, long = Technische Universit\"at Kaiserslautern}

\DeclareAcronym{PBCH}{short = PBCH , long = physical broadcast channel}
\DeclareAcronym{SS/PBCH}{short = SS/PBCH , long = synchronization signal / physical broadcast channel}
\DeclareAcronym{SS}{short = SS , long = synchronization signal}
\DeclareAcronym{SSS}{short = SSS , long = secondary synchronization signal}
\DeclareAcronym{PSS}{short = PSS , long = primary synchronization signal}
\DeclareAcronym{DMRS}{short = DMRS , long = demodulation reference signal}
\DeclareAcronym{PDSCH}{short = PDSCH , long = physical downlink shared channel}
\DeclareAcronym{PRACH}{short = PRACH , long = physical random access channel}
\DeclareAcronym{mMIMO}{short = mMIMO , long = massive multiple input multiple output}
\DeclareAcronym{SU-MIMO}{short = SU-MIMO , long = single-user multiple input multiple output}
\DeclareAcronym{MIMO}{short = MIMO , long = multiple input multiple output}
\DeclareAcronym{MU-MIMO}{short = MU-MIMO , long = multiple-user multiple input multiple output}
\DeclareAcronym{FWA}{short = FWA , long = fixed wireless access}

\DeclareAcronym{METAS}{short = METAS , long = Federal Institute of Metrology}
\DeclareAcronym{BNetzA}{short = BNetzA , long = {Bundesnetzagentur}}

\begin{document}

\title{Overview of the Evaluation Methods for the Maximum EMF Exposure in 5G Networks \\

\thanks{This work has been funded by the Federal Ministry for Digital and Transport of Germany (BMDV) (project number VB5GFKAISE). This is a preprint version, the full paper has been accepted by the IEEE Conference on Standards for Communications and Networking (CSCN), Thessaloniki, Greece. November 2022. Please cite as: A. Fellan and H. D. Schotten, “Overview of the evaluation methods for the maximum EMF exposure in 5G networks,” in IEEE Conference on Standards for Communications and Networking (CSCN), 2022}
}

\author{\IEEEauthorblockN{Amina Fellan\textsuperscript{$*$}, Hans D. Schotten\textsuperscript{$*$$\diamond$}}
\IEEEauthorblockA{\textsuperscript{$*$}\textit{Institute of Wireless Communication, Technische Universit\"{a}t Kaiserslautern,} Kaiserslautern, Germany.\\
	\{fellan, schotten\}@eit.uni-kl.de}
\textsuperscript{$\diamond$}\textit{Intelligent Networks, German Research Center for Artificial Intelligence (DFKI),} Kaiserslautern, Germany.\\ Hans$\_$Dieter.Schotten(at)dfki.de}

\maketitle

\begin{abstract}
% TO-DO: Amina, in progress
Instantaneous measurements of the \ac{EMF} strength do not reflect the maximum exposure levels possible in a given location. An extrapolation factor needs to be applied to the measurements before comparing them against the local exposure guidelines or recommendations for compliance evaluation. For the \ac{5G} networks, a standardized approach for extrapolating \ac{EMF} values is yet to be defined. This work provides an overview of the state-of-the-art research that focuses on estimating the maximum \ac{EMF} exposure caused by radiation from \ac{5G} base stations. It also considers current efforts by national and international organizations to establish standardized methods for extrapolating the \ac{EMF} measurements which is necessary in investigating conformance with the \ac{EMF} guidelines and regulations.    
\end{abstract}


\section{Introduction}
% TO-DO: Amina, Open

\begin{figure}[t]
\centerline{\includegraphics[width=1\columnwidth]{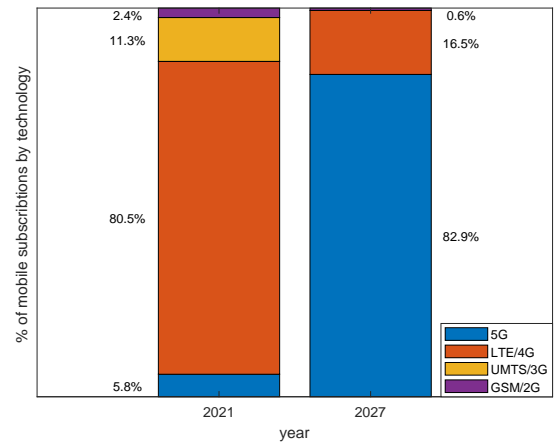}}
\caption{Recent and projected percentages of user subscriptions by mobile technology in Western Europe. Partially reproduced from \cite{Ericsson}.}
\label{fig_EU_mobile_tech}
\end{figure}

The first deployments of the \acf{5G} technology in Germany started in 2019. Since the start of \ac{5G} roll-out in 2019 with roughly 139 active \acp{BS} \cite{BNetzA_2021} and until this day, the \ac{5G} population coverage has rapidly increased from an estimated 18\% in 2020 to about 91\% in Q3 of 2022 \cite{5G_observatory_2022}.
At the time of this writing, it is projected that there are currently at least 65,905 active \ac{5G} \acp{BS} in operation; of which 15.1\% are assigned to the low-band (below 1 GHz), 17.9\% in the 3.6 GHz band and operating in \ac{SA} mode \cite{5G_observatory_2022}, and 67.1\% running in the \ac{DSS} mode in conjunction with \ac{LTE} technology in frequency bands n28, n3, and n1 corresponding to 700 MHz, 1.8 GHz and 2.1 GHz, respectively. The allocation of the high-band frequencies that are in the 26 GHz range has started in Germany and the results of the initial round of spectrum auctions have been concluded in May 2021 by awarding 5 campus network license grants in band n258; however, the list of awardees has not been disclosed to the public yet \cite{BNetzA_2020}. Only few \ac{EU} countries have started auctions on the n258 band (8 out of 27 at the time being) and even fewer have started testing services, some of the examples include the \ac{5G} \ac{mmWave} \ac{FWA} network by Fastweb in Italy \cite{5g_mmwave_freq_chiaraviglio} and \ac{mmWave} venue coverage by Nokia in Finland \cite{Nokia}.
Figure \ref{fig_EU_mobile_tech} presents a breakdown of the recent (from 2021) and projected user subscriptions based on mobile technologies \cite{Ericsson}. As can be seen from the figure, it is anticipated that by 2027 \ac{5G} will be the most dominant mobile technology in Western Europe (among other regions such as: North America, North East Asia, and Gulf Cooperation Council \cite{Ericsson}).

This rapid spread of \ac{5G} technology has sparked public concern on the increasing exposure of humans to the \acf{EMF} radiated by the emerging technology. Several organizations, such as the \ac{WHO}, \acf{ICNIRP}, \ac{IEEE}, and \ac{IEC}, work constantly to define and maintain guidelines and recommendations with the goal to protect humans and the environment against any possible harmful effects caused by the exposure to radio-frequency electromagnetic fields \cite{ICNIRP2020, IEEEstd2019, emf_fellan}. 

The \ac{EMF} exposure limits set by the regulatory organizations target the overall maximum possible radiation at a given frequency and location, which is not easily measurable in practice with regular measurement equipment as they only provide the instantaneous and average values of \ac{EMF} radiation at a given location. For proper comparison with the regulatory limits, extrapolation of the measured values to the maximum possible \ac{EMF} is necessary. This task was relatively straight-forward for previous generations of communication technologies \cite{environments7030022}. However, in \ac{5G} networks the task is more involved as several aspects and features that distinguish \ac{5G} from its predecessors need to be considered. 

This work investigates current research and standardization attempts for extrapolation of \ac{EMF} measurements within the context of \ac{5G} networks. Our work is structured as follows: in Section \ref{SoA}, we provide an overview of recent works that consider different extrapolation methods for \ac{EMF} measurements relating to \ac{5G} \ac{NR} technology. Section \ref{standards} outlines some of the national and international standardization endeavors for the assessment of \ac{EMF} exposure in \ac{5G}. Finally, the conclusions and outlooks for this work are given in Section \ref{conclusions}. 

% - what are EMF estimation methods based on ? which signals are measured ? 
% - short historical background
% - limits set by ICNIRP or national regulation correspond to the maximum EMF

\section{Assessment approaches: state-of-the-art}\label{SoA}

The general concept for approximating the maximum \ac{EMF} exposure caused by \ac{5G} is similar in principle to that used for previous communication generations. Namely, it is based on measuring the periodic signaling that is always transmitted and independent of a \ac{BS}'s traffic load \cite{environments7030022}. In the case of \ac{5G} \ac{NR}, the \ac{SS/PBCH} signaling block (or \ac{SSB} for short) in the downlink satisfies this requirement. 

However, there are some additional considerations that need to be taken into account for the \ac{EMF} evaluation in \ac{5G} networks. For instance, unlike with previous generations, the \ac{SSB} is not centered around the carrier frequency, nor does it have a fixed position in the \ac{5G} \ac{NR} frame. Instead, it can occur anywhere across the carrier bandwidth and thus needs to be carefully determined during measurements. Additionally, if beamforming is used, variations between the broadcast and traffic antennas patterns should be taken into consideration which could turn into quite the involved task \cite{5g_emf_mimo_adda}. Also, the transmission duty cycle of the \ac{TDD} mode, in terms of downlink to uplink ratio, needs to be reflected when evaluating the extrapolation factor. 

In this section, we examine some of the works from both the research and academia communities that studied the assessment of human exposure to \ac{EMF} and the problem of extrapolating the measured values to the maximum theoretically possible exposure within the context of \ac{5G} technology. The extrapolation methods for evaluating the maximum \ac{EMF} exposure can be generally categorized as follows: 
% TO-DO: Amina, Open
% main focus is works considering the maximum EMF (only)
\subsection{Based on frequency-selective methods}
Frequency-selective measurements employ a frequency-measuring equipment to evaluate \ac{EMF} emissions produced by a given technology. A properly configured spectrum analyzer, for instance, is capable of performing this type of measurement by evaluating the synchronisation signal's power in zero-span mode. Several works have based their proposed extrapolation techniques of \ac{EMF} exposure produced by \ac{5G} \acp{BS} on frequency-selective measurements. One of the earliest works addressing the extrapolation problem of \ac{EMF} in \ac{5G} from a theoretical perspective is \cite{Keller2019OnTA}, where Keller proposes two extrapolation methods; the first of which is based on frequency-selective measurements. He also defined the preconditions necessary for the application of this method. Authors in \cite{5G_DSS_schilling} applied an extrapolation factor for \ac{5G} base stations operating in \ac{NSA} mode that relies on determining the maximum number of subcarriers. The chosen method is an extension of the \ac{LTE} \ac{EMF} extrapolation method specified in \cite{IEC_62232_2017}, as \ac{5G} in \ac{DSS} mode shares the spectrum with \ac{LTE} technology and, for instance, uses the same \ac{SCS} of 15 KHz as \ac{LTE} to guarantee coexistence.  
An extrapolation factor that reflects the difference between the bandwidths of the full signal to that of the \ac{SSB} and between the maximal gains of the traffic and broadcast beams, is proposed by the authors in  \cite{5G_mMIMO_bornkessel}. 

In \cite{environments7030022}, \cite{5g_emf_mimo_adda,5G_PDSCH_adda, 5G_PDSCH_gnb_migliore}, the authors extend the extrapolation method to account for additional factors specific to \ac{5G} \ac{NR} by leveraging the frame structure of \ac{5G} to extract information on numerology and channel duty cycle, even when such information is not available from the operator. They also verify the proposal of including \ac{gNB} forced traffic beams with the help of dedicated \acp{UE} for \ac{PDSCH} measurements, and its associated \ac{DMRS}, which is vital for estimating the maximum \ac{EMF} accurately. 

Authors in \cite{5g_mmwave_freq_chiaraviglio} design and propose a measurement framework for a \ac{SA} \ac{mmWave} \ac{BS} scenario in a commercial deployment. They also consider what influence does user traffic have over the levels of exposure. 

However, this measurement method comes with its own limitations such as difficulty in distinguishing between downlink and uplink transmissions in \ac{TDD} mode, the lack of capacity to discern \ac{PBCH} from \ac{PRACH} signals \cite{IEC_62232_2017}, and the inability to differentiate between detected emissions from neighbouring beams of \ac{NR} cells. Consequently, frequency-selective methods might be appropriate for early stages of \ac{5G} deployments while bearing in mind their limitations. 

\subsection{Based on code-selective methods}
Code-selective measurements provide a higher level of insight and detail with respect to technology-specific \ac{EMF} emissions in comparison to frequency-selective methods; as the measurement equipment is capable of measuring and decoding the individual signaling components from the different mobile communication technologies. For \ac{5G} networks, it is able to deliver information distinguishing \acp{SSB} from different cells. 

\begin{table*}[ht]

\begin{center}
  \caption{Summary of recent works on the assessment of the maximum \ac{EMF} human exposure produced by \ac{5G} \acp{BS}}
  \label{tab:lit_comparison}
  \begin{tabular}{|c|c|c|c|c|c|l|}

    \hline
    \multirow{2}{*}{Ref.} & \multirow{2}{*}{publication date} & \multirow{2}{*}{\ac{5G} technology} & \multicolumn{2}{c|}{method}  & \multirow{2}{*}{measured signal} & \multirow{2}{*}{contribution} \\
    \cline{4-5}
    
            &  &  & {frequency-selective}  & {code-selective} &  &    \\
    \hline     
    \cite{Keller2019OnTA} & Nov 2019 & \acs{mMIMO} & X & X & \multicolumn{1}{m{2cm}|}{\ac{SSB} (\acs{PSS} / \acs{SSS})} & 
    \multicolumn{1}{m{5.2cm}|}{proposed two \ac{EMF} extrapolation methods for evaluating the maximum theoretically possible values and defined preconditions for their application} \\
    \hline
    \cite{5G_mMIMO_methodology_aerts} & Dec 2019 & \acs{mMIMO} & X &  & \ac{SSB} &
    \multicolumn{1}{m{5.2cm}|}{provided an experimental methodology for the assessment and extrapolation of \ac{EMF} exposure that considers gain difference between \ac{SSB} and traffic beams}\\
    \hline
    \cite{environments7030022} & Mar 2020 & \ac{mMIMO} & X & X  & \multicolumn{1}{m{2cm}|}{\ac{SSB} (\ac{PBCH}-\acs{DMRS}) \& \ac{PDSCH}}  &
    \multicolumn{1}{m{5.2cm}|}{proposed an \ac{EMF} extrapolation procedure for \ac{5G} signals and provided an experimental validation of it in a controlled environment} \\
    \hline
    \cite{5g_emf_mimo_adda} & May 2020 & \ac{mMIMO} & X &  X &  \multicolumn{1}{m{2cm}|}{\ac{SSB} (\ac{PBCH}-\acs{DMRS}) \& \ac{PDSCH} }&
    \multicolumn{1}{m{5.2cm}|}{provided a theoretical analysis of \ac{NR} signals at the physical layer level and experimentally verified the extrapolation of \ac{5G} \ac{EMF} exposure for a \ac{SU-MIMO} \ac{BS} scenario that can be generalized} \\
    \hline
    \cite{5G_mMIMO_bornkessel} & Mar 2021 & \ac{mMIMO} & X &   & \multicolumn{1}{m{2cm}|}{\ac{SSB} (\acs{PSS} / \ac{SSS})} &
    \multicolumn{1}{m{5.2cm}|}{proposed an extrapolation factor for frequency-selective measurements that takes into account the differences in the bandwidths and gains of the traffic and broadcast beams}\\
    \hline
    \cite{5G_mimo_bern_aerts} & Apr 2021 &  \ac{mMIMO} & X &   & \ac{SSB} \& \ac{PDSCH} &
    \multicolumn{1}{m{5.2cm}|}{suggested an extrapolation method that takes into account the influence of the traffic beam and that does not require additional information on the \ac{BS} from the operator} \\
    \hline
    \cite{5G_PDSCH_gnb_migliore} & Jun 2021 &  \ac{mMIMO} & X & X & \multicolumn{1}{m{2cm}|}{\ac{PDSCH} \& \ac{PDSCH}-\acs{DMRS}} &
    \multicolumn{1}{m{5.2cm}|}{investigated and compared different approaches of estimating the maximum power extrapolation based on active beam forcing with \acp{UE} and leveraging the \ac{5G} frame structure} \\
    \hline
    \cite{5g_mmwave_wali} & Jan 2022 & \acs{mmWave} &  & X & \ac{SSB} (\ac{SSS}) &
    \multicolumn{1}{m{5.2cm}|}{evaluated the maximum \ac{EMF} exposure from a \ac{5G} \ac{mmWave} \ac{BS} following the extrapolation procedure defined in \cite{IEC_62232_2017} } \\
    \hline
    \cite{5G_DSS_schilling} & Mar 2022 & \acs{DSS} & X & X &  \multicolumn{1}{m{2cm}|}{\ac{SSB} (\acs{PSS} / \ac{SSS})} &
    \multicolumn{1}{m{5.2cm}|}{proposed a novel code-selective \ac{EMF} extrapolation method for \ac{5G} \acp{BS} operating in \ac{DSS} mode and compared the results with those obtained using the frequency-selective method} \\
    \hline
    \cite{5G_DSS_adda} & Jul 2022 & \acs{DSS} &  & X & \ac{PBCH}-\acs{DMRS} & 
    \multicolumn{1}{m{5.2cm}|}{verified the applicability of the maximum power extrapolation methods to evaluate the maximum \ac{EMF} exposure from \ac{LTE} and \ac{5G} \acp{BS} operating in \ac{DSS} mode} \\
    \hline
    \cite{5g_mmwave_freq_chiaraviglio} & Aug 2022 & \acs{mmWave} & X &  & \ac{mmWave} carriers &
    \multicolumn{1}{m{5.2cm}|}{proposed a framework and a measurement algorithm for the evaluation of maximum \ac{EMF} exposure levels in a commercial \ac{5G} \ac{FWA} \ac{mmWave} deployment that does not require further extrapolation}\\
    \hline
    \cite{5g_mmwave_code_migliore} & Sep 2022 & \acs{mmWave} &  & X &  \ac{SSB} (\ac{SSS}) &
    \multicolumn{1}{m{5.2cm}|}{verified the applicability of the maximum power extrapolation method for a commercial \ac{5G} \ac{FWA} \ac{mmWave} \ac{BS} and performed an uncertainty budget analysis for their assessment methodology}\\
    
  \hline
  
\end{tabular}
\end{center}
\end{table*}

The second extrapolation method that Keller proposed in \cite{Keller2019OnTA} for the theoretical assessment of the max \ac{EMF} exposure is based on demodulation of the \ac{SS}. The exposure index is evaluated based on measurements of the \acf{PSS} and \acf{SSS} parts of the \ac{SSB}, which can be determined using a code-selective device. The authors in \cite{5g_mmwave_wali} extend the application of the methodology derived in \cite{5G_mMIMO_methodology_aerts} to evaluate the maximum exposure from a \ac{mmWave} \ac{BS} using a code-selective measurement setup and calculate the extrapolated \ac{EMF} using the formula adopted in \cite{IEC_62232_2017}. 
Authors in \cite{5G_DSS_schilling} and \cite{5G_DSS_adda} perform the \ac{EMF} exposure assessment using a code-selective method as well for both \ac{LTE} and \ac{5G} \acp{BS} operating in \ac{DSS} mode. They also highlight the superiority of code-selective measurements in comparison to frequency-selective ones for \ac{5G} deployments.

In Table \ref{tab:lit_comparison}, we provide a short summary of existing works addressing the extrapolation of \ac{EMF} measurements to evaluate maximum exposure in \ac{5G} networks. We highlight the following aspects about them: the extrapolation method used, type of \ac{5G} technology under study, and the measured signals used for evaluating the extrapolated values. The works are listed in a chronological order based on their publication dates.  

\section{Standardization efforts}\label{standards}
% TO-DO: Amina

In this section, we consider existing efforts on the national and international levels that provide guidelines for the extrapolation procedures of \ac{5G} \ac{EMF} measurements to the maximum theoretically possible levels.    

On the national level, the Federal Network Agency for Electricity, Gas, Telecommunications, Post and Railway, or \ac{BNetzA} for short, in Germany is concerned with setting the guidelines for the measurement and evaluation of the maximum \ac{EMF} exposure. Based on \cite{NRW}, the \ac{BNetzA} endorses two methods; both of which extrapolate the \ac{EMF} measurements using the frequency-selective method. The first method is based on beam forcing at maximum load to the point where the frequency-selective measurement is carried out. This method is limited to vendor-specific network equipment (due to the tool used for traffic forcing) and requires the support of the network operator. The second and more broadly applicable method is based on frequency-selective measurements of the \ac{SSB}, and more specifically, the \ac{SSS} part within a measurement bandwidth of 2 MHz. Extrapolation to the maximum \ac{EMF} takes into account the total bandwidth of the signal, a gain correction factor for \acp{BS} operating in beamforming mode which takes into account the difference between the traffic and broadcast beams, and a duty cycle correction factor for the operation in \ac{TDD} mode.

Another example of national efforts is provided by the \ac{METAS} in Switzerland in their technical report \cite{METAS_5GNR}. The \ac{METAS} defines two methods for extrapolation of \ac{EMF} for \ac{5G}  \ac{NR} \acp{BS} based on the previously described measurement approaches in Section \ref{SoA}. However, it adopts the code-selective method as the reference method whereas the frequency-selective method is seen merely as an approximate one since an overestimation is highly likely when using this method. Both extrapolation methods are based on the measurement of the \ac{SSS} part of the \ac{SSB}. In the code-selective method, the extrapolation is evaluated per \ac{NR} cell and the applied extrapolation factor takes into consideration an antenna correction factor that depends on the azimuth and elevation angles, a beam statistic factor that accounts for the use of adaptive antennae, and a duplex factor that depends on the downlink to uplink ratio for \ac{TDD} mode. The total extrapolated \ac{EMF} caused by all \ac{NR} cells is finally evaluated by applying the \ac{RSS} on all \ac{NR} cell-specific contributions. 
On the other hand, the frequency-selective extrapolation method can be executed by measuring the \ac{EMF} over the bandwidth of the \ac{SSS} using the max-hold function in a spectrum analyzer. It is then necessary to apply a reduction factor to adjust the measurement to reflect the value per \ac{RE} and a correction factor that accounts for installations supporting beamforming. 

On an international level, the \ac{IEC} published the IEC-62232 standard governing the methods of evaluation for human \ac{EMF} exposure using \ac{RF} measurements in the vicinity of mobile base stations \cite{IEC_62232_2017}. However, the latest edition of the standard that was published in 2017 addressed communication technologies up to \ac{LTE} and did not account for the unique features specific to the \ac{5G} \ac{NR} technology, as the first deployment of the latter started only in 2019 \cite{5G_observatory_2019}. An upgrade of the standard is currently undergoing finalization and is expected to be released and officially adopted by the end of 2022. The latest version of the \ac{IEC} standard will include extensions that are relevant to \ac{5G} \ac{NR} such as evaluation methods that consider \acp{BS} supporting \ac{mMIMO} and those operating in \ac{DSS} mode \cite{IEC62232_21}. It will also reflect modifications in the \ac{EMF} exposure limits adopted in the latest guidelines published by the \ac{ICNIRP} in 2020 \cite{ICNIRP2020}.

\section{Conclusions}\label{conclusions}
% TO-DO: Amina
In this work, we provided a short review of current research on the extrapolation of \ac{5G} \ac{NR} \ac{EMF} measurements to the maximum \ac{EMF} exposure values; a necessary step to ensure the compliance of \ac{5G} \ac{NR} \acp{BS} with the \ac{EMF} exposure limits and guidelines, set by organizations such as the \ac{ICNIRP} or the responsible national regulatory authorities, that are intended to protect humans and the environment. We classified the research efforts based on the measurement method used in the two main categories of frequency-selective and code-selective methods. We also summarized some of the national and international standardization endeavours that aim to set a reference method for evaluating the human \ac{EMF} exposure taking into consideration the recent \ac{5G} technology. 
We plan to use this overview as a foundation for choosing the appropriate measurement methodology and \ac{EMF} extrapolation method to assess our recently deployed \ac{5G} campus network in Kaiserslautern \cite{emf_fellan}.

% \begin{table}[htbp]
% \caption{Table Type Styles}
% \begin{center}
% \begin{tabular}{|c|c|c|c|}
% \hline
% \textbf{Table}&\multicolumn{3}{|c|}{\textbf{Table Column Head}} \\
% \cline{2-4} 
% \textbf{Head} & \textbf{\textit{Table column subhead}}& \textbf{\textit{Subhead}}& \textbf{\textit{Subhead}} \\
% \hline
% copy& More table copy$^{\mathrm{a}}$& &  \\
% \hline
% \multicolumn{4}{l}{$^{\mathrm{a}}$Sample of a Table footnote.}
% \end{tabular}
% \label{tab1}
% \end{center}
% \end{table}

% \begin{figure}[htbp]
% \centerline{\includegraphics{fig1.png}}
% \caption{Example of a figure caption.}
% \label{fig}
% \end{figure}

\section*{Acknowledgment}
This work is made possible through the funding provided by the Federal Ministry for Digital and Transport of Germany (BMDV) under the project “5x5G-Strategie” (project number VB5GFKAISE). The authors alone are responsible for the content of this paper.

\bibliography{References.bib} 
\bibliographystyle{ieeetr} 

\end{document}